\documentclass[epj]{svjour}

\usepackage[dvips]{graphicx}

\usepackage{array}
\usepackage{amssymb}
\usepackage{amsmath}
\usepackage{hhline}
\usepackage{longtable}
\usepackage{dcolumn}
\usepackage{bm}
\usepackage{subfigure}
\usepackage{epsfig}
\usepackage{latexsym,amsmath}
\usepackage{amsbsy}

\makeatletter
\renewcommand\appendix{\par
  \setcounter{section}{0}%
  \setcounter{subsection}{0}%
  \setcounter{equation}{0}%
  \renewcommand\theequation{\Alph{section}.\arabic{equation}}
  \renewcommand\thesection{Appendix \@Alph\c@section:}
  \renewcommand\thesubsection{\arabic{subsection}}}
\makeatother

\begin{document}
\title{Transition from fractal to non-fractal scalings in growing scale-free networks}

\author{Zhongzhi Zhang\inst{1,2} \thanks{e-mail: zhangzz@fudan.edu.cn} \and Shuigeng Zhou\inst{1,2} \thanks{e-mail: sgzhou@fudan.edu.cn} \and Lichao Chen\inst{1,2}  \and Jihong Guan\inst{3}}                     
\institute{Department of Computer Science and Engineering, Fudan
University, Shanghai 200433, China \and Shanghai Key Lab of
Intelligent Information Processing, Fudan University, Shanghai
200433, China \and Department of Computer Science and Technology,
Tongji University, 4800 Cao'an Road, Shanghai 201804, China}

\date{Received: date / Revised version: date}

\abstract{Real networks can be classified into two categories:
fractal networks and non-fractal networks. Here we introduce a
unifying model for the two types of networks. Our model network is
governed by a parameter $q$. We obtain 
the topological properties of the network including the
degree distribution, average path length, diameter, fractal
dimensions, and betweenness centrality distribution, which are
controlled by
parameter $q$. 
Interestingly, we show that by adjusting $q$, the networks undergo a
transition from fractal to non-fractal scalings, and exhibit a
crossover from `large' to small worlds at the same time. Our
research may shed some light on understanding the evolution and
relationships of fractal and non-fractal networks.  
} 

 \maketitle
\section{Introduction}

The past ten years have witnessed a considerable interest in
characterizing and understanding the topological properties of
networked systems~\cite{AlBa02,DoMe02,Ne03,BoLaMoChHw06,CoRoTrVi07}.
It has been established that small-world property~\cite{WaSt98} and
scale-free behavior~\cite{BaAl99} are the two most fundamental
concepts constituting our basic understanding of the organization of
many natural and social systems. A serial of recent research
indicate that these two features often go along, and have important
consequences on almost every aspect on dynamic processes taking
place on networks~\cite{Ne03,BoLaMoChHw06}. The small-world
characteristic means that the node number of network (order)
increases exponentially with the average path length (APL), and thus
leads to the general understanding that complex scale-free
small-world networks are not invariant or topologically fractal,
since fractal networks implies that there is a power-law relation
between the network order and its APL.

More recently, by using a renormalization procedure based on the
box-counting method, Song, Havlin and Makse discovered that some
real-life networks exhibit fractal scaling~\cite{SoHaMa05,SoHaMa06}.
The fractal topology can be characterized via two relevant
exponents: fractal dimension $d_B$ and degree exponent of the boxes
$d_k$. The fractal dimension $d_B$ is measured by the scaling of the
minimum number of boxes $N_{B}$ of linear size $\ell_{B}$ that is
needed to cover the network with order $N$, in other words,
$N_{B}/N\thicksim \ell_{B}^{-d_{B}}$. Similarly, one can identify
the degree exponent of the boxes through the relation
$k_{B}(\ell_{B})/ k_{hub} \thicksim \ell_{B}^{-d_{k}}$, where
$k_{B}(\ell_{B})$ is the degree of each node of the renormalized
network, and $k_{hub}$ the maximum degree of nodes inside each box
of original network. In fractal scale-free networks with degree
distribution $P(k)\sim k^{-\gamma}$, the three indexes $\gamma$,
$d_{B}$ and $d_{k}$ are related by $\gamma =1+d_{B}/
d_{k}$~\cite{SoHaMa05,SoHaMa06}.

According to the presence of fractal scaling or not, networks can be
assorted into two categories~\cite{SoHaMa05}: In the presence of
fractal behavior, a network is said to be fractal; in contrast, if a
network exhibits no fractal scaling, it is defined as non-fractal.
Examples of the first class of networks include the World Wide Web
(WWW), the actor collaboration networks, metabolic networks, and
yeast protein interaction networks. And instances of the second type
are the Internet, and most model networks such as the Barab\'asi
-Albert (BA) network~\cite{BaAl99}, the Watts-Strogatz (WS)
network~\cite{WaSt98}, and the Erd\"os-R\'enyi (ER) random
graph~\cite{ErRe60}. In addition to different topological
aspects~\cite{SoHaMa05,KiHaPaRiPaSt07}, the two types of networks
also have distinct consequences regarding the physics of dynamical
models running on
them~\cite{HiBe06,ZhZhZo07,ZhZhZoGu08,RoHaAv07,RoAv07,Hi07}.

Given the fact that real networks are either fractal or non-fractal,
it is consequently of fundamental importance to understand the
growth mechanisms and uncover the origins of different kinds of
networks. To this end, a wide variety of models have been
presented~\cite{AlBa02,DoMe02,BoLaMoChHw06}. However, previous
network models, to the best of our knowledge, can generate either
fractal networks or non-fractal ones, but rarely
both~\cite{SoHaMa06}. Thus, it seems quite natural and interesting
to set up a unifying framework looking for a deeper connection
between fractal and non-fractal networks. This is the purpose of the
current work.

In this paper, by introducing a simple network growth process we
propose a unifying scenario for fractal and non-fractal networks. We
analytically obtain many structural characteristics of the network,
including degree distribution, average path length, diameter,
fractal dimension, and betweenness centrality. The degree
distribution obeys a power law with an exponent varying
continuously. The obtained results on APL and diameter show that the
network undergoes a transition from a small- to large-world network.
More interestingly, the network exhibits a crossover behavior
between fractality and non-fractality.

\section{Network model}

This section is devoted to network construction and computation of
some related quantities.

\begin{figure}[h]
\begin{center}
\includegraphics[width=0.38\textwidth]{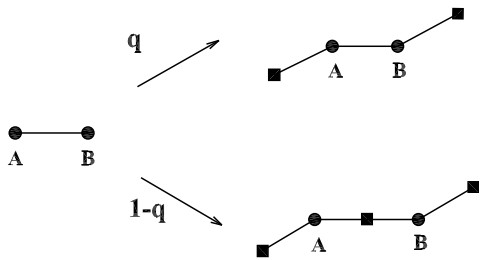}
\caption{Iterative construction method of the networks. Each link is
replaced by either of the two paths on the right-hand side of the
arrow with a certain probability, where each $\blacksquare$ stands
for a new node.}\label{fig1}
\end{center}
\end{figure}

\subsection{Construction algorithm}

The proposed evolving networks (graphs) have a treelike structure,
which are constructed in an iterative way as shown in
Fig.~\ref{fig1}. Let $T_{t}$ ($t\geq 0$) denote the networks after
$t$ iterations. For $t=0$, the networks growth begins from two nodes
(vertices) connected by an edge (link or bond). For $t\geq 1$,
$T_{t}$ is obtained from $T_{t-1}$. We replace each edge in
$T_{t-1}$ either by the path on the top right of Fig.~\ref{fig1}
with probability $q$, or by the path on bottom right with
complementary probability $1-q$. In other words, $T_t$ is generated
from $T_{t-1}$ by adding $k_v$ leaves to each node $v$, where $k_v$
is the degree of node $v$ in $T_{t-1}$. Then, the resulting graph is
further modified by expanding each edge with probability $1-q$,
which was already present in $T_{t-1}$. Expansion of an edge $(uv)$
means: removal of $(uv)$ and creation of an additional node $w$ with
edges $(uw)$ and $(vw)$. (Note that real systems may exhibit such an
evolving mechanism. For example, this mechanism has been used to
model the network evolution of connected minima on a potential
energy landscape~\cite{MaDo07}.) This procedure is iterated $t$
times, with the needed graphs obtained in the limit $t \to \infty$.

As will be shown in next section, when $q=1$, the network is a small
world with infinite fractal dimension, its average path length (APL)
grows logarithmically with node number. In the particular case of
$q=0$, the network is a `large' world with a finite fractal
dimension of 2, its APL scales exponentially with network size.
Except these two limiting cases of $q=1$ and $q=0$, for other $q$
($0<q<1$), the networks are growing stochastically. Varying $q$ in
the interval (0,1), the networks exhibit a transition from small to
large world, and simultaneously show an interesting phenomenon with
a transition from non-fractal to fractal behavior.

Note that the current model is similar to the probabilistic model
introduced by Song, Havlin, and Makse (SHM Model)
in~\cite{SoHaMa06}. The construction step with probability $q$ in
the current manuscript is analogous to the Mode I growth process in
the SHM Model, involving strong hub-hub attraction. And the step
with probability $1-q$ is analogous to the Mode II growth process in
the SHM Model, involving hub-hub repulsion. The probabilistic mixing
of these two types of construction steps leads to the same results
in both of the SHM Model and the current model: when $q=1$ (or pure
Mode I) is involved, the network has infinite fractal dimension and
shows small-world scaling; any $q \leq 1$ (or mixing of Modes I and
II) leads to a finite fractal dimension. However, in~\cite{SoHaMa06}
only part of the properties was addressed. Here we will present an
exhaustive analysis of various characteristics, including additional
features (such as average path length and betweenness centrality)
not calculated in~\cite{SoHaMa06}.

\subsection{Order and size}

Now we compute some related quantities such as the number of total
nodes and edges in $T_t$, called network order and size,
respectively. It should be mentioned that as $q$ is a real number,
we assume that all variables concerned with $q$ change continuously.
Note that a similar assumption was used in
Refs.~\cite{AlBa02,DoMe02,Ne03,BoLaMoChHw06}, which is valid for
large $t$. Let $L_v(t)$ be the number of nodes generated at step
$t$, $E_t$ the total number of edges present at step $t$. Then
$L_v(0)=2$ and $E_0=1$. By construction (see Fig.~\ref{fig1}), at
each time step, each existing edge is replaced either by three edges
with probability $q$ or by four edges with complementary probability
$1-q$. Thus, $E_t=q\times3E_{t-1}+(1-q)\times4E_{t-1}=(4-q)^t$
($t\geq 0$). At the same time, each existing edge yields two or
three new nodes with probability $q$ or $1-q$, this leads to
$L_v(t)=q\times2E_{t-1}+(1-q)\times3E_{t-1}=(3-q)(4-q)^{t-1}$
($t\geq 1$). Then the number of total nodes $N_t$ present at step
$t$ is
\begin{equation}\label{Nt1}
N_t=\sum_{t_i=0}^{t}L_v(t_i)=(4-q)^t+1.
\end{equation}
The average node degree after $t$ iterations is $\langle k
 \rangle_t=\frac{2\,E_t}{N_t}$,
which approaches $2$ for large $t$, coinciding with the treelike
structure of networks.

\section{Topological properties}
In this section, we will show that the tunable parameter $q$ in the
above construction algorithm controls the relevant features of the
networks.

\subsection{Degree distribution}
When a new node $i$ is added to the networks at a certain step $t_i$
($t_i\geq 1$), it has a degree of either 1 or 2. We denote by
$k_i(t)$ the degree of node $i$ at time $t$. By construction, the
degree $k_i(t)$ evolves with time as $k_i(t)=2\,k_i(t-1)$. That is
to say, the degree of node $i$ increases by a factor 2 at each time
step. Thus, the degree spectrum of the networks is discrete. In
network $T_t$ all possible degree of nodes is 1, 2, $2^2$ $2^3$,
$\ldots$, $2^{t-1}$, $2^t$.

Let $L_1(t_i)$ and $L_2(t_i)$ be the separate number of new nodes
with degree 1 and 2 that were born at step $t_i$. According to the
construction algorithm, we have
$L_1(t_i)=q\times2E_{t_i-1}+(1-q)\times2E_{t_i-1}=2\,(4-q)^{t_i-1}$
and $L_2(t_i)=(1-q)E_{t_i-1}=(1-q)(4-q)^{t_i-1}$. Then in network
$T_t$, the expected number of nodes of degree $k=2^{t-m}$ is
$n_k=L_1(m)+L_2(m+1)=(6-5q+q^2)(4-q)^{m-1}$.

Since the degree spectrum of the networks is not continuous. It
follows that the cumulative degree distribution~\cite{Ne03} is given
by $P_{\rm cum}(k)=\frac{N_{t,k}}{N_t}$, where $N_{t,k}=\sum_{k'\geq
k}n_{k'}$ is the number of nodes whose degree is not less than $k$.
When $t$ is large enough, we find $P_{\rm cum}(k)\approx
k^{-\ln(4-q)/\ln2}$. So the degree distribution $P(k)$ of the
networks follows a power-law form $P(k)\sim k^{-\gamma}$ with the
exponent
\begin{equation}\label{gamma}
\gamma=1+\frac{\ln(4-q)}{\ln2},
\end{equation}
which is a monotonically decreasing function of $q$. As $q$
increases from 0 to 1, $\gamma$ drops form 2 to
$1+\frac{\ln3}{\ln2}$~\cite{JuKiKa02}.

\subsection{Average path length}
Shortest paths play an important role both in the transport and
communication within a network and in the characterization of the
internal structure of the network~\cite{CoRoTrVi07}. Let $d_{ij}$
represent the shortest path length from node $i$ to $j$, then the
average path length (APL) $d_{t}$ of $T_t$ is defined as the mean of
$d_{ij}$ over all couples of nodes in the network, and the maximum
value $D_{t}$ of $d_{ij}$ is called the diameter of the network. APL
is relevant in many fields regarding real-life networks and has
received much attention~\cite{FrFrHo04}.

For general $q$, it is difficult to derive a closed formula for the
APL $d_{t}$ of network $T_t$. But for two limiting cases of $q=0$
and $q=1$, both the networks are deterministic ones, which allows
one to obtain the analytic solutions for APL. The detailed exact
derivation about APL is included in the Appendix section. The
obtained results show that the APL for these two particular cases
presents qualitatively disparate behaviors: For $q=0$, it is found
that
\begin{equation}\label{APL1}
d_t = \frac{8+7\times4^t+ 13\times8^t}{14 \left(1+4^t\right)},
\end{equation}
which is approximately equal to $\frac{13\times2^t}{14}$ for large
$t$. Since $N_t \sim 4^t$ for large network, so $d_t \sim
N_t^{1/2}$, indicating that $d_{t}$ grows as a square power of the
network order $N_t$. This phenomenon is similar to that of the
two-dimensional regular lattice~\cite{Ne00}. Thus, the network
corresponding to $q=0$ is not a small world. For $q=1$, we find
\begin{equation}\label{APL2}
d_t = \frac{2 \left(1+2\times3^t+ t\times3^t\right)}{3
\left(1+3^t\right)},
\end{equation}
which approximates $\frac{2\,t}{3}$ in the infinite $t$, implying
that the APL shows a logarithmic scaling with network order.
Therefore, in the specific case of $q=1$, the network exhibits a
small-world behavior.

Thus, when we tune $q$ from 0 to 1, the networks undergo a
transition from a `large'  to small world. We stress that such a
transition has already been observed in some previously studied
models~\cite{KlEg02}.

\subsection{Diameter}

As mentioned in preceding subsection, the diameter of a network is
defined as the longest shortest path between all pairs of nodes,
characterizing the maximum communication delay in the network.
Although we do not give a closed formula of APL of $T_t$ for general
$q$, here we will provide the exact result of the diameter of $T_t$
denoted by $D_t$.

We first address the network of $q=1$ case, where the shortest
distances between existing node pairs are not altered when new nodes
enter the systems. For this particular case, $D_0=1$. At each time
step, the diameter of the network increases by 2. Then the diameter
of $T_t$ is $D_t=2\,t+1$ and thus scales logarithmically with the
network order, showing a similar behavior as that of the average
path length. Since small diameter is consistent with the concept of
`small-world', the additive growth in the diameter with time also
(as the APL) suggests that the network for $q=1$ case is a small
world.

For $0\leq q<1$, the addition of new nodes affects fundamentally the
distances between existing node pairs. By construction algorithm,
for any existing couple of nodes connected by a link, after a
generation of evolution, the distant $l$ between this pair of nodes
may be equal to 1 or 2 with probability $q$ and $1-q$, respectively.
Thus the expected value of $l$ is $2-q$. Using this result and
considering that the networks are treelike, we can derive the
following recursive relation for expected $D_t$:
\begin{equation}\label{Dt1}
D_t = (2-q)D_{t-1}+2.
\end{equation}
Since $D_0=1$, we can resolve Eq.~\eqref{Dt1} to obtain the average
of network diameter as
\begin{equation}\label{Dt2}
D_t = \left(1+\frac{2}{1-q}\right)(2-q)^t-\frac{2}{1-q},
\end{equation}
which grows as a power of time $t$.

\subsection{Fractal dimension}

To determine the fractal dimension, we distinguish two cases: $0\leq
q<1$ and $q=1$. In the case of $0\leq q<1$, we follow the
mathematical framework introduced in Ref.~\cite{SoHaMa06}. We are
concerned about three quantities: network order $N_t$, network
diameter $D_t$, and degree $k_u(t)$ of a given node $u$. By
construction, we can easily see that in the infinite $t$ limit,
these quantities grow obeying the following relations: $N_t\simeq
(4-q)\,N_{t-1}$, $D_{t}\simeq (2-q)\,D_{t-1}$, $k_u(t)=
2\,k_u(t-1)$. Thus, for large networks, $N_t$, $k_u(t)$ and $D_t$
increase by a factor of $f_N=4-q$, $f_k=2$, and $f_D=2-q$,
respectively.

From above obtained microscopic parameters demonstrating the
mechanism for network growth, we can derive the scaling exponents:
the fractal dimension $d_B =\frac{\ln f_N}{\ln
f_D}=\frac{\ln(4-q)}{\ln(2-q)}$ and the degree exponent of boxes
$d_k =\frac{\ln f_k}{\ln f_D}=\frac{\ln2}{\ln(2-q)}$. According to
the scaling relation of fractal scale-free networks, the exponent of
the degree distribution satisfies $\gamma
=1+\frac{d_B}{d_k}=1+\frac{\ln(4-q)}{\ln2}$, giving the same
$\gamma$ as that obtained in the direct calculation of the degree
distribution, see Eq.~\eqref{gamma}.

For $q=1$, although the number of their nodes increases
exponentially, its diameter grows linearly with time. Thus, in this
case the network has infinite dimension and does not present a
fractal topology.

\subsection{Betweenness centrality}

Betweenness centrality (BC) of a given node is the accumulated
fraction of the total number of shortest paths going through the
node over all node pairs~\cite{Fr77,Newman01}. More precisely, the
betweenness of a node $i$ is
\begin{equation}\label{Between1}
b_{i}=\sum_{j \ne i \neq k}\frac{\sigma_{jk}(i)}{\sigma_{jk}},
\end{equation}
where $\sigma_{jk}$ is the total number of shortest path between
node $j$ and $k$, and $\sigma_{jk}(i)$ is the number of shortest
path running through node $i$.

We now investigate the BC distribution of nodes. It was claimed in
previous studies that for scale-free networks, the BC distribution
$P(b)$ of nodes obeys a power law with the exponent $\gamma_b=2$,
i.e., $P(b)\sim b^{-2}$. It was also suggested that the exponent
$\gamma_b=2$ is universal for all scale-free
networks~\cite{GoKaKi01}. Next, we will present that the exponent
$\gamma_b$ is not universal and constant (at least for fractal
trees), but varies significantly as a function of $d_B$ (or a
function or $q$).

In order to obtain the exponent $\gamma_b$ of BC distribution of our
networks, we resort to a heuristic argument similar to that applied
in~\cite{BrWuChBuKaSrcoLoHaSt07}. Notice that all of our networks
are treelike and fractal (except the case of $q=1$). Considering one
network with dimension $d_B$, for a small area of the network
containing $g$ nodes, its average path length is typically
$d(g)=g^{1/d_B}$~\cite{BuHa96}. All nodes in this small region can
reach the rest nodes in the network via $d(g)$ nodes. Thus, the BC
of these $d(g)$ nodes is not less than $g$. On the other hand, in
the whole network there are $N_t/g$ such areas, each of which
includes $g$ nodes. Then, the total number of nodes with BC $b$ not
less than $g$ in the whole network is
\begin{equation}\label{Between2}
n(b\geq g) \sim d(g)\times \frac{N_t}{g}\sim N_t\times
g^{-(1-1/d_B)}.
\end{equation}
Therefore, the cumulative BC distribution is
\begin{equation}\label{Between3}
P_{cum}(b)=\frac{n(b\geq g)}{N_t} \sim  g^{-(1-1/d_B)},
\end{equation}
which implies $P(b)\sim b^{-\gamma_b}$ with
\begin{equation}\label{Between3}
\gamma_b=2-\frac{1}{d_B}.
\end{equation}
Thus, $\gamma_b$ is a increasing function of fractal dimension
$d_B$, the larger the fractal dimension $d_B$, the larger the
exponent $\gamma_b$. Eq.~\eqref{Between3} shows that $\gamma_b=2$
holds only for the non-fractal networks with $d_B\rightarrow
\infty$~\cite{GhOhGoKaKi04}, where a relatively small number of hub
nodes bear large BC. In contrast, in fractal networks, the BC of a
lot of `small' nodes can be compared with that of hub nodes.

\section{Conclusions}

In the paper, by introducing a parameter $q$, we have presented a
simple network growth process to generate a unified model for
fractal and non-fractal networks. This process was shown to lead to
a rich behavior for the network structure. Various relevant
topological properties have been determined depending on the model
parameter $q$. It has been shown that both degree and betweenness
centrality distributions of nodes have a power-law tail and the
characteristic exponents change continuously with the parameter $q$.
Other structural properties including the APL, diameter, and fractal
dimension have been obtained as well, which indicate that the model
undergoes a crossover from large-world to small-world networks, and
simultaneously exhibits a transition from fractal to non-fractal
behaviors.

In spite of the simplicity, our minimal model can capture the
essential characteristics and correlations of fractal and
non-fractal networks. It is helpful for understanding the growth
mechanisms and evolutions of the two different sorts of networks.
Finally, it should be mentioned that, although we only studied
treelike networks, in a similar way, one can construct models
considering the effect of loops, whose general properties are
similar to those of the model investigated in the present work.

\section*{Acknowledgment}
We thank Yichao Zhang for preparing this manuscript. This research
was supported by the National Basic Research Program of China under
grant No. 2007CB310806, the National Natural Science Foundation of
China under Grant Nos. 60496327, 60573183, 60773123, and 60704044,
the Shanghai Natural Science Foundation under Grant No. 06ZR14013,
the Postdoctoral Science Foundation of China under Grant No.
20060400162, Shanghai Leading Academic Discipline Project No. B114,
the Program for New Century Excellent Talents in University of China
(NCET-06-0376), and the Huawei Foundation of Science and Technology
(YJCB2007031IN).

\appendix

\section{Derivation of the average path length for two limited cases}

Following an algebraic method similar to that introduced
in~\cite{HiBe06}, we can compute the average path length (APL) for
two limiting deterministic cases. By definition, the APL for $T_t$
is defined as
\begin{equation}\label{eq:app1}
  d_{t}  = \frac{S_t}{N_t(N_t-1)/2}\,,
\end{equation}
where
\begin{equation}\label{eq:app2}
  S_t = \sum_{i \in T_{t}, j \in T_{t}, i\neq j} d_{ij}
\end{equation}
denotes the sum of the shortest path length between two nodes over
all pairs. For the two particular cases $q=0$ and $q=1$, Both of the
networks have a self-similar structure that allows one to calculate
$d_{t}$ analytically. The self-similar structure is obvious from an
equivalent network construction method: to obtain $T_{t+1}$, one can
make some copies of $T_{t}$ and join them in the hub nodes.

\subsection{The $q=0$ case}

\begin{figure*}
\centering
\includegraphics[width=0.70\textwidth]{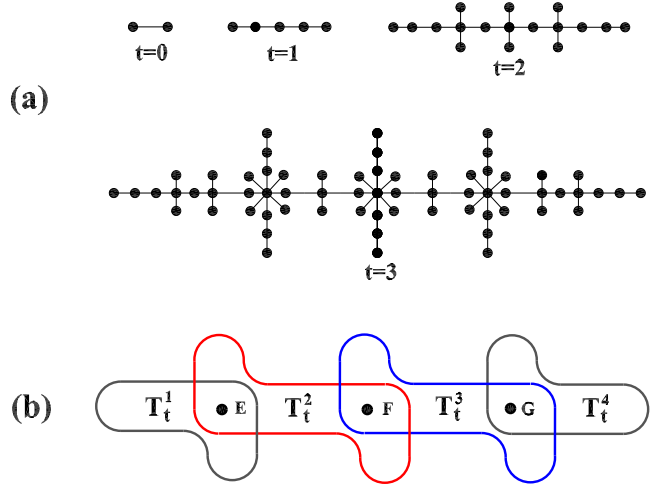}
\caption[kurzform] {(Color online) The network growth process for
the particular case $q=0$. (a) Illustration of the first four
evolution steps. (b) Second construction method of the network for
$q=0$ case that highlights self-similarity: The graph after $t+1$
    construction steps, $T_{t+1}$, is composed of four copies of
    $T_t$ denoted as
    $T_t^{\theta}$ $(\theta=1,2,3,4)$, which are
    connected to one another as above. }\label {apfig2}
\end{figure*}

As shown in Fig.~\ref{apfig2}, for the  $q=0$ case, the network
$T_{t+1}$ may be obtained by the juxtaposition of four copies of
$T_t$, which are labeled as $T_{t}^{\theta}$, $\theta=1,2,3,4$. Then
we can write the sum $S_{t+1}$ as
\begin{equation}\label{eq:app3}
  S_{t+1} = 4\,S_t + \Delta_t\,,
\end{equation}
where $\Delta_t$ is the sum over all shortest paths whose endpoints
are not in the same $F_{t}$ branch. The solution of
Eq.~\eqref{eq:app3} is
\begin{equation}\label{eq:app4}
  S_t = 4^{t-1} S_1 + \sum_{x=1}^{t-1} 4^{t-x-1} \Delta_x\,.
\end{equation}
The paths that contribute to $\Delta_t$ must all go through at least
one of the three edge nodes (i.e., $\textbf{\emph{E}}$,
$\textbf{\emph{F}}$ and $\textbf{\emph{G}}$ in Fig.~\ref{apfig2}(b))
at which the different $T_t$ branches are connected. The analytical
expression for $\Delta_t$, called the crossing paths, is found
below.

Denote $\Delta_t^{\alpha,\beta}$ as the sum of all shortest paths
with endpoints in $T_t^{\alpha}$ and $T_t^{\beta}$. If
$T_t^{\alpha}$ and $T_t^{\beta}$ meet at an edge node,
$\Delta_t^{\alpha,\beta}$ rules out the paths where either endpoint
is that shared edge node. If $T_t^{\alpha}$ and $T_t^{\beta}$ do not
meet, $\Delta_t^{\alpha,\beta}$ excludes the paths where either
endpoint is any edge node.  Then the total sum $\Delta_t$ is
\begin{align}
\Delta_t =& \,\Delta_t^{1,2} + \Delta_t^{1,3} + \Delta_t^{1,4}+
\Delta_t^{2,3} + \Delta_t^{2,4}+\Delta_n^{3,4}. \label{eq:app5}
\end{align}

By symmetry, $\Delta_n^{1,2} = \Delta_t^{2,3} = \Delta_n^{3,4}$ and
$\Delta_t^{1,3} = \Delta_t^{2,4}$, so that
\begin{equation}\label{eq:app6}
\Delta_t = 3 \Delta_t^{1,2} + 2\Delta_t^{1,3} + \Delta_t^{1,4}\,.
\end{equation}
In order to find $\Delta_t^{1,2}$, $\Delta_t^{1,3}$, and
$\Delta_t^{1,4}$, we define
\begin{align}
s_t = \sum_{i \in T_t^{1},i\ne E}d_{i,E}\,. \label{eq:app7}
\end{align}
Considering the self-similar network structure, we can easily know
that at time $t+1$, the quantity $s_{t+1}$ evolves recursively as
\begin{eqnarray}
s_{t+1}
&=&2\,s_t+\left[s_t+2^t(N_t-1)\right]+\left[s_t+2^{t+1}(N_t-1)\right]\nonumber\\
&=&4\,s_t+3\times2^{3t}.\label{eq:app8}
\end{eqnarray}
Using $s_1=7$, we have
\begin{eqnarray}
s_t=2^{2t-2}+3\times2^{3t-2}.
\end{eqnarray}
On the other hand, by definition given above, we have
\begin{eqnarray}
  \Delta_t^{1,2} &=& \sum_{\substack{i \in T_t^{1},\,\,j\in
      T_t^{2}\\ i,j \ne E}} d_{ij}\nonumber\\
  &=& \sum_{\substack{i \in T_t^{1},\,\,j\in
      T_t^{2}\\ i,j \ne E}} (d_{iE} + d_{jE}) \nonumber\\
  &=& (N_t-1)\sum_{\substack{i \in T_t^{1}\\ i \ne E}} d_{iE} + (N_t-1) \sum_{\substack{j \in T_t^{2}\\ j \ne E}} d_{jE} \nonumber\\
  &=& 2(N_t-1)\sum_{i \in T_t^{1}\,i \ne E} d_{iE}\nonumber\\
  &=& 2(N_t-1)\,s_t,
\label{eq:app9}
\end{eqnarray}
\begin{eqnarray}
  \Delta_t^{1,3} &=& \sum_{\substack{i \in T_t^{1},\,i \ne E\\ j\in
      T_t^{3},\,j \ne F}} d_{ij}\nonumber\\
  &=& \sum_{\substack{i \in T_t^{1},\,i \ne E\\ j\in
      T_t^{3},\,j \ne F}} (d_{iE} + d_{EF}+ d_{jF}) \nonumber\\
  &=& 2(N_t-1)\,s_t+(N_t-1)^2\times 2^t\,,
\label{eq:app10}
\end{eqnarray}
and
\begin{eqnarray}
  \Delta_t^{1,4} &=& \sum_{\substack{i \in T_t^{1},\,i \ne E\\ j\in
      T_t^{4},\,j \ne G}} d_{ij}\nonumber\\
  &=& \sum_{\substack{i \in T_t^{1},\,i \ne E\\ j\in
      T_t^{4},\,j \ne G}} (d_{iE} + d_{EG}+ d_{jG}) \nonumber\\
  &=& 2(N_t-1)\,s_t+(N_t-1)^2\times2^{t+1}\,,
\label{eq:app11}
\end{eqnarray}
where $d_{EF}=2^t$ and $d_{EG}=2^{t+1}$ have been used. Substituting
Eqs.~(\ref{eq:app9}), (\ref{eq:app10}) and (\ref{eq:app11}) into
Eq.~(\ref{eq:app6}), we obtain
\begin{eqnarray}\label{eq:app12}
\Delta_t &=& 12(N_t-1)\,s_t+4\,(N_t-1)^2\times2^{t}\,\nonumber\\
&=&13\times32^{t}+3\times16^{t}.
\end{eqnarray}
Inserting Eqs.~(\ref{eq:app12}) for $\Delta_x$ into
Eq.~(\ref{eq:app4}), and using $S_1 = 20$, we have
\begin{equation}
 S_t = \frac{4^t}{28}\times \left(8+7\times4^t+ 13\times8^t\right). \label{eq:app13}
\end{equation}
Inserting Eq.~\eqref{eq:app13} into Eq.~\eqref{eq:app1}, one can
obtain the analytical expression for $d_t$ in Eq.~(\ref{APL1}).

\subsection{The $q=1$ case}
\begin{figure*}
\begin{center}
\includegraphics[width=0.70\textwidth]{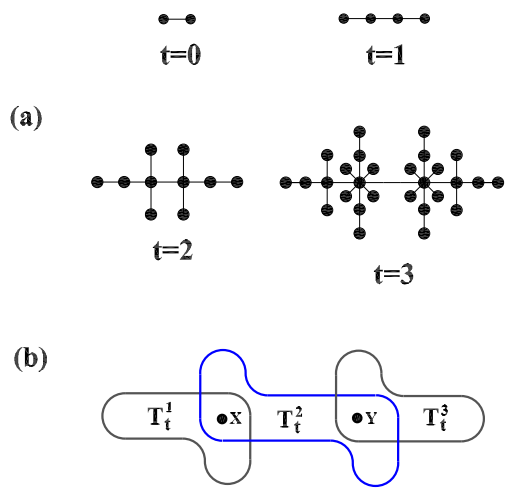}
\caption[kurzform] {(Color online) The network growth process for
the specific case $q=1$. (a) Scheme of the first four iterative
processes. (b) Second construction means of the network for $q=1$
case:  The graph after $t+1$ construction steps, $T_{t+1}$, can be
obtained by joining three copies of $T_t$ denoted as $T_t^{\psi}$
$(\psi=1,2,3)$, which are connected to each other at the two edge
nodes ($X$ and $Y$).}\label {apfig3}
\end{center}
\end{figure*}

Using analogous analysis for the $q=0$ case, we can calculate the
APL $d_t$ for the case of $q=1$. For simplicity, we use the same
symbols used in last subsection to represent the identical notions.
For $q=1$ case, the network $T_{t+1}$ may be obtained by joining at
the hubs (the most connected nodes) three copies of $T_t$ labeled as
$T_t^{(\psi)}$, $\psi=1,2,3$~\cite{Bobe05}, see Fig.~\ref{apfig3}.
Then one can write the sum over all shortest paths $S_{t+1}$ as
\begin{equation}\label{eq:app14}
  S_{t+1} = 3\,S_t + \Delta_t\,.
\end{equation}
The solution of Eq.~(\ref{eq:app14}) is
\begin{equation}\label{eq:app15}
  S_t = 3^{t-1} S_1 + \sum_{\tau=1}^{t-1} 3^{t-\tau-1} \Delta_\tau\,.
\end{equation}

The paths contributing to $\Delta_t$ must all go through at least
either of the two hubs ($X$ and $Y$) where the three different $T_t$
branches are joined. The crossing paths $\Delta_t$ is given by
\begin{equation}\label{eq:app16}
\Delta_t=\Delta_t^{1,2} + \Delta_t^{2,3} + \Delta_t^{1,3}\,,
\end{equation}
where $\Delta_t^{1,2}$ rules out the paths where either endpoint is
node $X$, $\Delta_t^{2,3}$ rules out the paths where either endpoint
is node $Y$, and $\Delta_t^{1,3}$ excludes the paths with an
endpoint is either $X$ or $Y$. Again by symmetry, $\Delta_t^{1,2} =
\Delta_t^{2,3}$, so that
\begin{equation}\label{eq:app17}
\Delta_t = 2 \Delta_t^{1,2}+\Delta_t^{1,3}\,.
\end{equation}
In this case, the quantity $s_{t+1}$ evolves as
\begin{eqnarray}
s_{t+1}
&=&2\,s_t+[s_t+(N_t-1)]\nonumber\\
&=&3\,s_t+3^{t}.\label{eq:app18}
\end{eqnarray}
 Since
$s_1 = 4$, Eq.~\eqref{eq:app18} is solved inductively:
\begin{equation}
s_t = (t+3)\times3^{t-1}\,. \label{eq:app19}
\end{equation}
similarly,
\begin{eqnarray}
  \Delta_t^{1,2} &=& \sum_{\substack{i \in T_t^{1},\,\,j\in
      T_t^{2}\\ i,j \ne X}} (d_{iX} + d_{jX}) \nonumber\\
   &=& 2(N_t-1)\,s_t,
\label{eq:app20}
\end{eqnarray}
and
\begin{eqnarray}
  \Delta_t^{1,3}  &=& \sum_{\substack{i \in T_t^{1},\,i \ne X\\ j\in
      T_t^{3},\,j \ne Y}} (d_{iX} + d_{XY}+ d_{jY}) \nonumber\\
  &=& 2(N_t-1)\,s_t+(N_t-1)^2\,.
\label{eq:app21}
\end{eqnarray}
Substituting the obtained expressions in Eqs.~\eqref{eq:app20} and
\eqref{eq:app21} into Eq.~\eqref{eq:app17}, the crossing paths
$\Delta_t$ is found to be
\begin{equation}\label{eq:app22}
  \Delta_t = 7\times 9^{t}+2t\times9^{t}.
\end{equation}
Inserting Eq.~\eqref{eq:app22} into Eq.~\eqref{eq:app15} and using
the initial condition $S_1 = 10$, we have
\begin{equation}\label{APL11}
  S_t = 3^{-1+t} \left(1+2\times3^t+ t\times3^t\right).
\end{equation}
Substituting Eq.~(\ref{APL11}) into (\ref{eq:app1}), the exact
expression for the average path length is obtained as shown in
Eq.~(\ref{APL2}).


\end{document}